\documentstyle[11pt,paspconf,epsf]{article}

\begin{document}
\title{NICMOS Imaging of High-Redshift Radio Galaxies}
\author{A.~Zirm~(JHU),
A.~Dey~(NOAO),
M.~Dickinson~(STScI),
P.~J.~McCarthy~(OCIW),
P.~Eisenhardt~(JPL/Caltech),
S.~G.~Djorgovski~(Caltech),
H.~Spinrad~(UCB),\\
S.~A.~Stanford and W.~van~Breugel~(IGPP/LLNL)}

\begin{abstract}
We have obtained near--infrared (1.6 $\mu$m) images of 11 powerful 3CR 
radio galaxies at redshifts $0.8 < z < 1.8$ using NICMOS on board HST. 
The high angular resolution permits a detailed study of galaxy 
morphology in these systems at rest--frame optical wavelengths, 
where starlight dominates over the extended, aligned UV 
continuum.   The NICMOS morphologies are mostly symmetric and are 
consistent with dynamically relaxed, elliptical host galaxies dominated 
by a red, mature stellar population.  The aligned structures are sometimes 
faintly visible, and nuclear point sources may be present in a few 
cases which manifest the ``unveiled'' AGN that is obscured from view 
at optical wavelengths.  Our observations are consistent with the 
hypothesis that the host galaxies of $z\approx 1-2$ radio galaxies are 
similar to modern--day gE galaxies.  Their sizes are typical of gE 
galaxies but smaller than present--day cD and brightest cluster 
galaxies, and their surface brightnesses are higher, as expected given 
simple luminosity evolution.  
  
\end{abstract}

\keywords{galaxies: active}

\section{Introduction}
At one time, radio sources offered the only readily available
means of locating galaxies at large redshifts ($z > 1$), and were
thus studied as a window on the early history of galaxy evolution.  
Radio galaxies obey a tight near--infrared Hubble ($K$--$z$)
relation and are frequently associated with rich cluster environments.
This suggests that there is an evolutionary sequence linking high-redshift
radio galaxies to low-redshift giant ellipticals and cD galaxies.
The discovery that many high-redshift radio galaxies have elongated, 
complex UV continuum structures aligned with the radio source
axis (McCarthy et al.\ 1987; Chambers et al.\ 1987) suggested that the
active nucleus might affect the UV morphology and possibly even 
the evolutionary history of the host galaxy.  It was believed that 
some radio galaxies might be true ``protogalaxies'' forming the bulk 
of their stars via some process induced by the radio jets.  
However, later studies have shown that in many cases the aligned 
UV continuum arises largely from scattered AGN emission 
(Di Serego Alighieri et al.\ 1989) and/or nebular continuum 
emission (Dickson et al.\ 1995).

The spectacular, complex structures seen in optical WFPC2 images of 
3CR radio galaxies by (e.g.) Best et al.\ (1997) are, in many 
cases, AGN--related ``ephemera'' surrounding a more normal host galaxy.
However, the brightness of this aligned, blue light makes it difficult to 
study the properties of the underlying stellar component when observing 
at optical wavelengths.  Observers have therefore turned to the 
near--IR, where the AGN--related emission is fainter and the stellar 
component brighter.  Rigler et al.\ (1992), Dickinson et al.\ (1994)
and Best et al.\ (1998), among others, have shown that some 3CR radio 
galaxies are rounder and more symmetric when observed in the near--IR, 
suggesting the presence of relatively ``normal'' host galaxies underlying 
the UV--bright, aligned continuum (but see also Eisenhardt \& Chokshi 1990).
But until now these ground--based studies have been limited by angular 
resolution.  Now, using NICMOS on board HST, we can for the first time 
study the near--IR morphologies of high redshift galaxies with resolution 
comparable to that of the pioneering WFPC2 studies of these same objects.

\begin{figure}
\plotfiddle{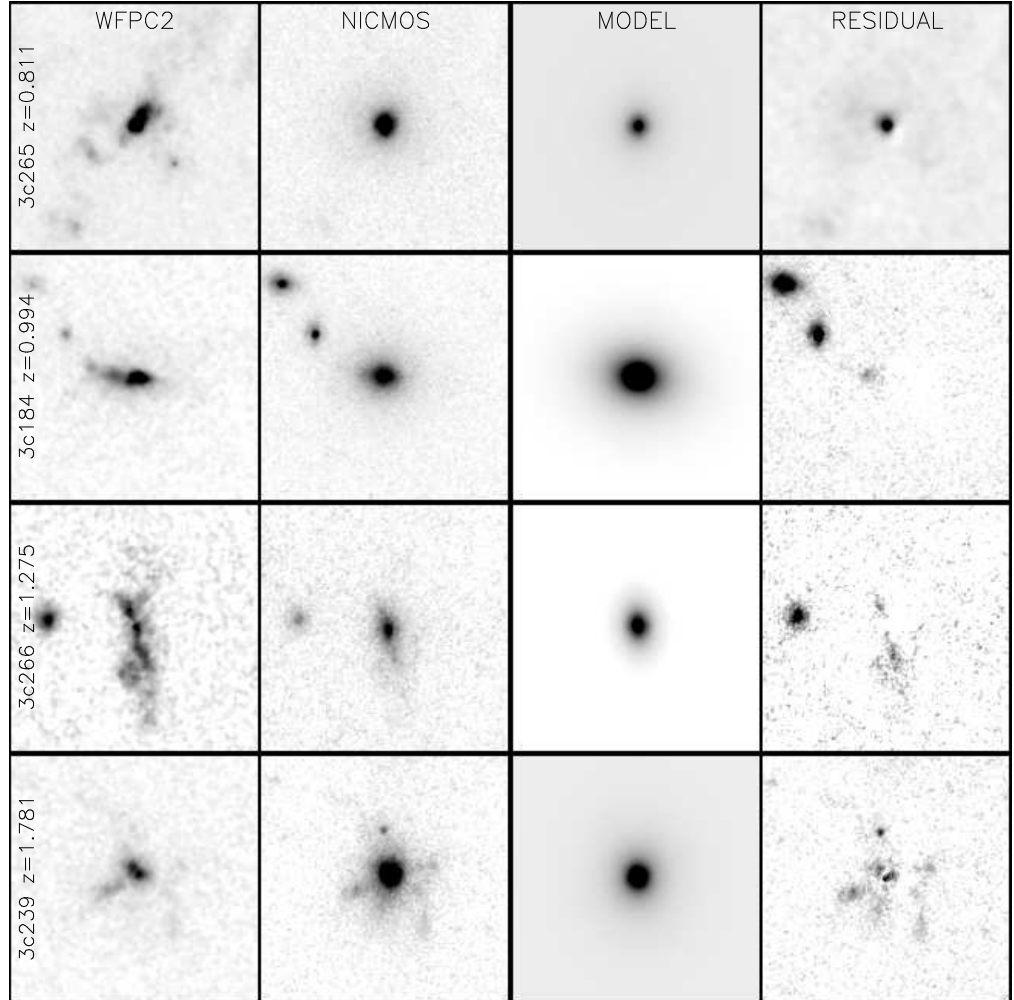}{4.9in}{0}{75}{75}{-205}{-25} 
\caption{WFPC2 and NICMOS images of four galaxies from our NICMOS imaging
sample, along with best--fitting models to the NICMOS host galaixes
and the NICMOS minus model residuals.}
\end{figure}

\section{Observations }

Our sample consists of 11 3CR radio galaxies at $0.8 < z < 1.8$,
imaged with NICMOS Camera 2, which provides diffraction limited
images at 1.6$\mu$m.  We used bandpasses (F160W or F165M) which
avoid strong nebular emission lines, which could significantly 
contaminate the fluxes and affect the observed morphologies.
Optical WFPC2 imaging is available for the whole sample:  in many 
cases we have unusually deep and often polarimetric WFPC2 data, 
while in others archival data by Best et al.\ 1997 or other sources 
were used.   In most cases we also have extensive ground--based 
supporting data (spectroscopy, polarimetry, etc.) from the 
W.M. Keck Observatory and other facilities.  

\begin{figure}
\plotone{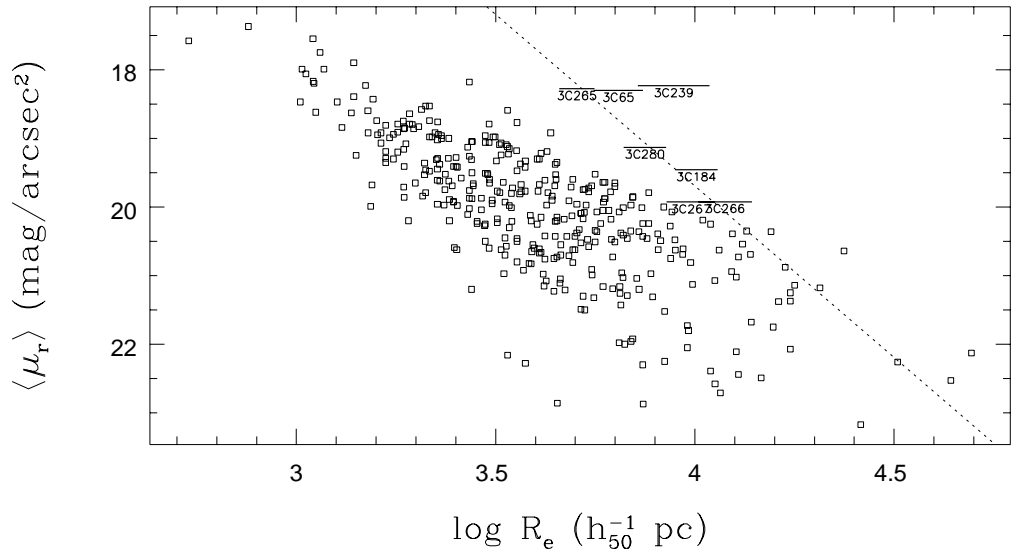} 
\caption{Rest frame Gunn $r$ size--surface brightness relation for 
local cluster ellipticals (small squares; J{\o}rgensen et al. 1995) 
and for our NICMOS radio galaxies (lines).  The lines connect the 
effective radii for q$_{0}$ = 0 and q$_{0}$ = 0.5 cosmologies.  
The dotted line shows the relation for constant galaxy luminosity, 
as expected for ``standard candle'' galaxies.}
\end{figure}

\section{Discussion}

The NICMOS and WFPC2 images of four galaxies from this sample are shown 
in Figure 1.   We have fit PSF--convolved models to the NICMOS images 
using a hybrid scheme which matches 1D surface brightness profiles and 
2D PA~+~ellipticity information.  The models and the NICMOS image 
residuals after model subtraction are also shown in Figure 1.
In most cases, the NICMOS images show that the rest--frame optical
light from powerful 3CR radio galaxies at $z > 1$ is rounder,
smoother, more symmetric and centrally concentrated than that observed
at rest--frame UV wavelengths.  The complex, aligned structures seen
in WFPC2 images are generally much less pronounced in the near--IR,
although in several cases (e.g. 3C 280, 3C 266, 3C 368) the highest
surface brightness regions of the aligned components can still be
detected.  In several cases, the near--IR surface brightness peaks
at the position of a local {\it minimum} in the WFPC2 images,
suggesting the effects of dust lanes affecting the near--UV 
morphologies.   A few galaxies (e.g., 3C~265) appear to have nuclear 
point sources in the IR, possibly showing the ``unveiled'' AGN.

Overall, the gross morphologies and surface brightness profiles 
of most 3CR hosts are consistent with their being high luminosity 
giant elliptical galaxies, already structurally mature.  This may be 
true as early as $z = 1.8$, although at that redshift 3C~239 appears 
to have ``ragged edges'' perhaps suggesting that it is in the process 
of accreting material through mergers.   However, the most distant 
galaxy in our sample, 3C~256 at $z = 1.82$, is radically different
than the others.  It is elongated, aligned, diffuse, and underluminous, 
and thus may be the exceptional example of a young radio galaxy early 
in the stages of its formation (see also Eisenhardt \& Dickinson 1992,
Simpson et al.\ 1999).

In Figure 2, we plot surface brightness vs.\ effective radius
(the ``Kormendy relation'') for 6 galaxies which are well fit
by $R^{1/4}$--law models, converting the NICMOS photometry 
(rest--frame $\lambda_0$0.57 to 0.88$\mu$m for our sample)
to rest--frame Gunn~$r$ ($\lambda_0 0.65 \mu$m) for comparison
to nearby cluster ellipticals.  The galaxies are physically smaller 
than the largest and brightest giant cluster ellipticals at $z = 0$,
and have higher rest--frame surface brightnesses as would be
expected given nominal luminosity evolution.  Most fall on the 
locus of constant luminosity (see also Best et al.\ 1998),
as might be expected given the small $K$--$z$ scatter.  
3C~239 at $z=1.78$ is significantly more luminous for 
its size compared to the galaxies at $0.8 < z < 1.3$.

\acknowledgments
 
Support for this work was provided by NASA grant GO--07454.02--96A.
Thanks to Hy for getting us all into this radio galaxy mess!

\end{document}